\documentclass[aps,twocolumn,epsf,showpacs,floatfix]{revtex4}
\usepackage{amssymb}
\usepackage{subfigure}
\usepackage{graphicx}
\usepackage[usenames]{color}

\begin{document}

\title{Gap solitons in superfluid boson-fermion mixtures}
\author{Sadhan K. Adhikari}
\thanks{Electronic address: adhikari@ift.unesp.br; URL:
http://www.ift.unesp.br/users/adhikari}
\affiliation{Instituto de F\'{\i}sica Te\'{o}rica, UNESP -- S\~{a}o Paulo State
University, 01.405-900 S\~{a}o Paulo, S\~{a}o Paulo, Brazil }
\author{Boris A. Malomed}
\thanks{Electronic address: malomed@eng.tau.ac.il; URL:
http://www.eng.tau.ac.il/~malomed/}
\affiliation{Department of Physical Electronics, School of Electrical Engineering,\\
Faculty of Engineering, Tel Aviv University, Tel Aviv 69978, Israel}

\begin{abstract}
Using coupled equations for the bosonic and fermionic order parameters, we
construct families of gap solitons (GSs) in a nearly one-dimensional
Bose-Fermi mixture trapped in a periodic optical-lattice (OL) potential,
the boson and fermion components being in the states of the BEC\ and BCS
superfluid, respectively.  Fundamental GSs are compact states trapped,
essentially, in a single cell of the lattice. Full families of such
solutions are constructed in the first two bandgaps of the OL-induced
spectrum, by means of variational and numerical methods, which are found to
be in good agreement. The families include both \textit{intra-gap} and
\textit{inter-gap} solitons, with the chemical potentials of the boson and
fermion components falling in the same or different bandgaps, 
respectively.
Nonfundamental states, extended over several lattice cells, are constructed
too. The GSs are stable against strong perturbations.
\end{abstract}

\maketitle

\section{Introduction}

The experimental realization of Bose-Einstein condensation (BEC) \cite{books}
was followed by the creation of another ultra-cold low-density quantum
medium, \textit{viz.}, the degenerate Fermi gas (DFG)\ \cite{Jin}. Fermi
gases cannot be directly produced by means of the evaporative cooling, due
to the Pauli blockade of interactions among spin-polarized fermions.
However, it has been possible to achieve the DFG state using the
sympathetic-cooling technique, in the presence of another boson or fermion
component in the gas. The use of this technique has resulted in the
observation \cite{Jin,exp2,exp3,exp4} and related experimental \cite%
{exp5,exp5x,exp6} and theoretical \cite{yyy,zzz,capu,ska} studies of trapped
DFG-BEC mixtures composed of the following species: $^{7}$Li-$^{6}$Li \cite%
{exp3}, $^{23}$Na-$^{6}$Li \cite{exp4}, and $^{87}$Rb-$^{40}$K \cite%
{exp5,exp5x}. New remarkable experiments have made use of the
Feshbach-resonance tool to induce attraction between fermions in spin-up and
spin-down states, thus creating the Bardeen-Cooper-Schrieffer (BCS)
superfluid \cite{bcs} in Fermi gases of $^{6}$Li \cite{grimm} and $^{40}$K
\cite{djin} atoms. Actually, the attraction in these experiments was varied
from weak, appropriate to the BCS state per se, to strong, appropriate to
the formation of the BEC of diatomic molecules, making it possible to
explicitly study the BCS-BEC crossover, as theoretically predicted in
earlier works \cite{cross} and analyzed in more recent ones \cite{cross2} (a
review of experimental results can be found in Ref. \cite{Jin-review}, 
and a
brief overview of the theory was given in Ref. \cite{C}).

Among most fundamental matter-wave patterns observed in BEC are
bright solitons, see original works \cite{Li-soliton,Rb-soliton}
and reviews \cite{BECsolitons}. Further analysis has predicted the
formation of bright solitons in degenerate Bose-Fermi
\cite{BFsoliton(BBrepBFattr),Sadhan-BFsoliton} and Fermi-Fermi
mixtures \cite{ff}, in the presence of attractive inter-species
interactions, which may overcome the effective Pauli repulsion
between identical fermions (and possible repulsion in the BEC
component).

On the other hand, \textit{gap solitons} (GSs) have been predicted
\cite{GSprediction} and further studied
\cite{stable,Fatkhulla,Alfimov,Arik,Thawatchai,Beata} in BEC with
the repulsion between atoms, trapped in the periodic potential
induced by an optical lattice (OL). GSs are localized objects
supported by the balance between the self-repulsion and negative
effective mass provided by the OL-induced matter-wave spectrum.
The chemical potential of the GS must fall in a spectrum's
bandgap, where delocalized Bloch-wave states cannot exist.

The prediction of GSs in the BEC \cite{GSprediction} was followed
by the creation of a GS formed by  $\simeq 250$ atoms of
$^{87}$Rb in a ``cigar-shaped" trap combined with the longitudinal
OL potential \cite{Markus,Markus-review}. In the experiment, the
BEC was pushed in a state with the negative effective mass by
means of acceleration. Another possibility may be to add a strong
parabolic trap to the OL potential, thus squeezing the system into
a small region and then slowly relaxing the extra trap, to give (a
part of) the condensate a chance to remain in a squeezed GS state
\cite{Michal}. Nonfundamental states, in the form of a stretched
``gap-wave" pattern, extending over many cells of the OL, have
also been observed \cite{gap-wave} and interpreted theoretically
as a segment of a nonlinear Bloch wave confined by sharp fronts
\cite{gap-wave-theory} (in fact, similar states were also
predicted in the self-attractive BEC \cite{attractive}).

Multicomponent GSs were predicted too. A straightforward
possibility is to consider a symmetric model for a mixture of two
boson species with equal masses and coefficients of the coupling
to the OL potential, which may represent two different hyperfine
states of the same atom, with repulsion between the species. As
shown in Ref. \cite{Arik}, this model gives rise to stable 2D and
1D gap solitons -- first of all, in the most fundamental case when
the intra-species repulsion is absent (as may be adjusted by means
of the Feshbach-resonance technique), and the GSs are supported,
as symbiotic \cite{sym} states, solely by the inter-species
interaction. These two-component GSs may be of \textit{intra-gap}
and \textit{inter-gap} types, with the chemical potentials of the
components falling into a single bandgap, or belonging to
different gaps, respectively (inter-gap solitons are essentially
less stable \cite{Arik}). Also studied were three-component GSs in
the 1D model of a spinor BEC\ trapped in the OL \cite{Beata}.

In this work we aim to study the possibility of the formation of 
fundamental
GSs in a one-dimensional (1D) model of the intrinsically repulsive
superfluid Bose-Fermi mixture (SBFM), whose boson and fermion components 
are
assumed to be in the BEC and BCS-superfluid phases, respectively.
Fundamental GSs, unless taken very close to edges of the corresponding
bandgap, feature a compact shape, being essentially localized in a single
well of the OL potential, without tangible undulating tails. Obviously, a
localized pattern cannot be a ground state of the gas with repulsive
interactions. Nevertheless, the GSs in intrinsically self-repulsive BEC have
been found to be stable against small or moderate perturbations \cite{stable}
(i.e., they are, generally speaking, metastable states), and, as mentioned
above, they have been created in  experiment 
\cite{Markus,Markus-review}.
We also show that, in addition to the compact fundamental GSs, one can find
solutions for \textit{nonfundamental} solitons, which occupy several OL
sites and are dynamically stable too.

It is well known that a dilute BEC is very accurately described by
Gross-Pitaevskii equation (GPE) for the mean-field wave function, which, in
particular, provides for a reliable model for the dynamics of dark and
bright solitons, vortices, onset of collapse, etc. \cite{books,BECsolitons}.
In the Bose-Fermi mixture, the bosonic component is also treated by means of
the GPE (with an extra term taking into regard collisions with fermions),
whereas the fermionic part should be, in principle, described by a set of
Schr\"{o}dinger equations for individual atoms \cite{Mario}. However, with
the increase of the number of fermions such \textit{ab initio} treatment
becomes much too complicated to implement. On the other hand, a simplified
model of the BCS superfluid of the Ginzburg-Landau type is widely accepted
\cite{bcs}. This approach describes the fermion superfluid by a single
complex order parameter, which may be interpreted, at a semi-microscopic
level, as the wave function of composite bosons (Cooper pairs). The model
based on nonlinearly coupled equations for the boson and fermion order
parameters has been used to study various phenomena in Bose-Fermi mixtures,
such as vortex formation, mixing-demixing transitions, and the onset of
collapse \cite{capu,SadhanMixDemix,bfvortex}. We will use these equations
for the prediction of compact GSs in the SBFM, which seems quite a 
relevant
approach to the description of these simple-shaped objects.

The so derived model for the SBFM in three dimensions amounts to a
system of nonlinear Schr\"{o}dinger equations, with the cubic
nonlinearity for the bosons, and nonlinear term of power 7/3 for
the fermions \cite{skcol,Sadhan-collapse}; the two equations are
coupled by cubic terms. In the presence of a strong transverse
trap, this system can be reduced to coupled 1D equations. We look
for GS solutions and analyze their stability within the framework
of the 1D system. The simplicity of the Gaussian-like profiles of
the GSs suggests to apply the variational approximation (VA)
\cite{VA} to them, which produces very accurate results, if
compared to numerical findings.

It is relevant to mention that states of the GS type in a Bose-Fermi 
mixture
were also investigated in Ref. \cite{Mario}, using a microscopic description
of the fermion component. Unlike the compact GSs studied in the present
work, the solitons found in Ref. \cite{Mario} typically occupy several OL
cells, and feature a sophisticated shape with undulating tails. Another
difference is that Ref. \cite{Mario} was chiefly dealing with the attractive
Bose-Fermi interaction, whereas in our system both the Bose-Fermi and
Bose-Bose interactions could be  repulsive as well as attractive.

The paper is organized as follows. In Sec. II we derive the basic equations
from the underlying Lagrangian density. Section III is dealing with the VA
based on the Gaussian ansatz for the wave functions. In Sec. IV we report
numerical results for intra-gap and inter-gap GS families in the first two
bandgaps, and compare them to the VA predictions, concluding that the
agreement between the numerical and variational methods is very good. We
also verify the stability of the GSs by means of direct simulations. Section
V summarizes the work.

\section{Coupled\ equations for the superfluid Bose-Fermi mixture 
(SBFM)}

\subsection{Three-dimensional equations}

We consider a degenerate mixture of $N_{B}$ bosons and $N_{F}$ fermions with
spin $\frac{1}{2}$, whose masses are $m_{B}$ and $m_{F}$, at zero
temperature. The boson and fermion components of the mixture are assumed to
be condensed into the BEC and a BCS superfluid, respectively. Accordingly, a
weakly attractive underlying (hidden) interaction between spin-up and
spin-down fermions is implied, while the explicit fermion-boson and
boson-boson interactions are repulsive, which is the most natural case for
any atomic gas.

The derivation of the model starts with the effective Lagrangian density for
the SBFM,
\begin{equation}
\mathcal{L}=\mathcal{L}_{B}+\mathcal{L}_{F}+\mathcal{L}_{BF},  \label{yy}
\end{equation}%
with the usual expression for the boson component \cite{books},
\begin{eqnarray}
\mathcal{L}_{B} &=&\frac{i\hbar }{2}\left( \psi _{B}\frac{\partial \psi
_{B}^{\ast }}{\partial t}-\psi _{B}^{\ast }\frac{\partial \psi _{B}}{%
\partial t}\right) +\frac{\hbar ^{2}}{2m_{B}}|\nabla \psi _{B}|^{2}  \nonumber \\
&+&V_{B}(\mathbf{r})|\psi _{B}|^{2} +
\frac{1}{2}G_{B}|\psi _{B}|^{4},  \label{LB}
\end{eqnarray}%
where $\psi _{B}$ is the mean-field order parameter (single-atom wave
function) of the BEC, the density of the boson atoms being $n_{B}=|\psi
_{B}|^{2}$, and $V_{B}(\mathbf{r})$ is the trapping potential for the
bosons. The strength of the boson-boson repulsive interaction is $G_{B}=4\pi
\hbar ^{2}a_{BB}/m_{B}$, with $a_{BB}>0$ the respective scattering length.

To derive the Lagrangian density of the fermion component, we use the known
energy density of the BCS superfluid (SF)\ \cite{yang1,yang2,salasnich},
\begin{equation}
\mathcal{E}_{\mathrm{SF}}=(3/5){n}_{F}\varepsilon _{F},  \label{1A}
\end{equation}%
where $\varepsilon _{F}=\hbar ^{2}k_{F}^{2}/(2m)$ is the Fermi energy, $%
\hbar k_{F}$ is the Fermi momentum, and ${n}_{F}$ the density of the fermion
atoms. This energy density was first derived by Lee and Yang \cite{yang2} in
the weak-coupling BCS limit. Modifications to this expression for the
description of the BCS-BEC crossover (with the gradual increase of the
strength of the attraction between spin-up and spin-down fermions) have also
been considered \cite{salasnich}. With regard to the pairing of the fermions
with opposite spin orientations, the total fermion density is given by ${n}%
_{F}=2{(2\pi )^{-3}}]\int_{0}^{k_{F}}4\pi k^{2}dk\equiv \varepsilon
_{F}^{3/2}/A ^{3/2}$, with $A \equiv (3\pi ^{2})^{2/3}\hbar
^{2}/(2m_{F})$, which yields the expression for the Fermi energy in terms of
the density: $\varepsilon _{F}=A ({n}_{F}/2)^{2/3}$. Then, the energy
density in Eq. (\ref{1A}) is expressed as $\mathcal{E}_{\mathrm{SF}%
}=(3/5)A {n}_{F}^{5/3},$ and the respective fermionic Lagrangian
density in Eq. (\ref{yy}) becomes
\begin{eqnarray}
\mathcal{L}_{F} &=&\frac{i\hbar }{2}
\left( \psi _{F}\frac{\partial \psi
_{F}^{\ast }}{\partial t}-\psi _{F}^{\ast }\frac{\partial 
\psi _{F}}{%
\partial t}\right) +
\frac{\hbar ^{2}}{2\tilde m_{F}}|\nabla \psi _{F}|^{2}
  \nonumber \\&+&V_{F}(\mathbf{r})|\psi 
_{F}|^{2}
+\frac{3}{5}A |\psi _{F}|^{10/3},  \label{LF}
\end{eqnarray}%
where $\psi _{F}$ is the above-mentioned complex order parameter of the
fermionic BCS condensate, which determines its density, $n_{F}=|\psi
_{F}|^{2}$,  $V_{F}(\mathbf{r})$ is the potential trapping the fermion
atoms, and {$\tilde m_F$ is an effective mass of the 
superfluid flow; (it 
is natural to expect that in a weakly coupled superfluid $\tilde m_F$ 
is close to $2m_F$.)}

Lastly, the Bose-Fermi repulsive interaction is accounted for by the
corresponding term in the Lagrangian density,
\begin{equation}
\mathcal{L}_{BF}=G_{BF}|\psi _{B}|^{2}|\psi _{F}|^{2},  \label{LBF}
\end{equation}%
where $G_{BF}=2\pi \hbar ^{2}a_{BF}/m_{R}$, with the respective reduced
mass, $m_{R}=m_{B}m_{F}/(m_{B}+m_{F})$, and scattering length $a_{BF}$.
Substituting expressions (\ref{LB}), (\ref{LF}), and (\ref{LBF}) in full
Lagrangian density (\ref{yy}), we derive the following Euler-Lagrange
equations, cf. Ref. \cite{ska}:
\begin{eqnarray}
&&\left[ -i\hbar \frac{\partial }{\partial t}-\frac{\hbar ^{2}}{2m_{{B}}}%
\nabla_{\mathbf{r}} ^{2}+V_{{B}}(\mathbf{r})+G_{{B}}\left\vert \psi 
_{B}\right\vert
^{2}\right.   \nonumber \\
&&\left. +G_{{BF}}|\psi_F|^2\right] \psi _{{B}}(\mathbf{r},t)=0,  
\label{B} \\
&&\left[ -i\hbar \frac{\partial }{\partial t}-\frac{\hbar ^{2}\nabla _{%
\mathbf{r}}^{2}}{2\tilde m_F}+V_{{F}}(\mathbf{r})+A \left\vert \psi
_{F}\right\vert ^{4/3}\right.   \nonumber \\
&&\left. +G_{{BF}}\left\vert \psi _{B}\right\vert ^{2}\right] \psi _{F}(%
\mathbf{r},t)=0,  \label{F}
\end{eqnarray}%
which are supplemented with normalizations
\begin{equation}
\int \int \int \left\vert \psi _{B,F}(x,y,z)\right\vert ^{2}dxdydz=N_{B,F}
\label{NBF}
\end{equation}%
[these two norms are dynamical invariants of Eqs. (\ref{B}) and (\ref{F})].

{
Dynamical equations (\ref{B}) and (\ref{F}) can also be derived in a
different physical context, viz., in a DFG in the
hydrodynamic limit \cite{EPL}.  Indeed, using the known energy density
for the DFG, and a kinetic energy term
$\hbar^2/(2m_F)|\nabla \psi_F|^2$ for the hydrodynamic flow in the
Lagrangian density (here, $m_F$ is the effective mass of the 
hydrodynamic flow in the DFG), one arrives at Eqs. 
(\ref{B}) and
(\ref{F}) with the difference that the term $A=(3\pi^2
)^{2/3}\hbar^2/(2m_F)$ therein is replaced by $(6\pi^2
)^{2/3}\hbar^2/(2m_F)$; the difference in the factor of 2 is due to the 
formation of Cooper pairs in the superfluid. Hence the present analysis 
and conclusions 
for the stationary GSs should  also apply  to the DFG. However, this 
analogy 
may be questionable for nonstationary states, as the above-mentioned 
description of the DFG does not define the phase of the fermion order 
parameter.}

\subsection{Reduction to one-dimensional equations}

As said above, the presence of a strongly elongated (cigar-shaped) trap
suggest to reduce the 3D equations to a 1D system. In the usual GPE with the
cubic nonlinearity, the reduction of the full 3D equation to its 1D
counterpart was performed, in different ways, in Refs. \cite{Luca,CQ}. In
the simplest situation, the reduction of the 3D equation to one dimension
starts with the factorization of the 3D wave function,
\begin{equation}
\psi _{B,F}(x,y,z,t)=\phi _{B,F}(x,t)\exp \left( -i\left( \omega _{\perp
}\right) _{B,F}t-\frac{y^{2}+z^{2}}{2\left( a_{\mathrm{ho}}\right) _{B,F}^{2}%
}\right) ,  \label{psiphi}
\end{equation}%
where the second multiplier represents the ground state of the transverse 2D
harmonic oscillator, with $\left( a_{\mathrm{ho}}\right) _{B,F}=\sqrt{\hbar /%
\left[ m_{B,F}\left( \omega _{\perp }\right) _{B,F}\right] }$ the
harmonic-oscillator lengths for the bosons and fermions, and $\left( \omega
_{\perp }\right) _{B,F}$ the corresponding transverse trap frequencies. The
normalization of functions $\phi _{B,F}(x)$, which were introduced in Eq. (%
\ref{psiphi}), must comply with the underlying 3D normalization conditions (%
\ref{NBF}). The substitution of ansatz (\ref{psiphi}) in Eqs. (\ref{B}) and (%
\ref{F}) and averaging of the 3D equation in the transverse plane, $\left(
y,z\right) $, lead to the following 1D equations:
\begin{eqnarray}
i\hbar \frac{\partial \phi _{B}}{\partial t} &=&-\frac{\hbar ^{2}}{2m_{B}}%
\frac{\partial ^{2}\phi _{B}}{\partial x^{2}}+\frac{1}{2}G_{B}|\phi
_{B}|^{2}\phi _{B}  \nonumber \\
&+&\frac{1}{2}G_{BF}|\phi _{F}|^{2}\phi _{B}-\epsilon _{B}\cos \left( \frac{%
4\pi }{\lambda }x\right) \phi _{B},  \label{g} \\
i\hbar \frac{\partial \phi _{F}}{\partial t} &=&-\frac{\hbar 
^{2}}{2\tilde m_F}%
\frac{\partial ^{2}\phi _{F}}{\partial x^{2}}+\frac{3}{5}A |\phi
_{F}|^{4/3}\phi _{F}  \nonumber \\
&+&\frac{1}{2}G_{BF}|\phi _{B}|^{2}\phi _{F}-\epsilon _{F}\cos \left( \frac{%
4\pi }{\lambda }x\right) \phi _{F},  \label{h}
\end{eqnarray}%
where we have introduced the periodic OL potential with period $\lambda /2$
and amplitudes $-\epsilon _{B,F}$, acting on the Bose and Fermi components
(the amplitudes are taken negative to set a local minimum of the potential
at point $x=0$, where the center of the solitons will be placed). Below,
results are presented assuming equal boson and fermion masses  
 {and equal constants of the coupling to the OL: 
$\tilde m_F=m_{B}\equiv m$}, $\epsilon _{B}=\epsilon _{F}\equiv \epsilon $. 
{
For a Bose-Fermi superfluid mixture  
this situation may
approximately correspond to the existing $^{40}K$-$^{87}$Rb mixture
\cite{Jin}, where $\tilde m_F\approx 2m_F \approx m_{B}$}.

By means of the rescalings (here we set $m=m_{F}=m_{B}$ and $\epsilon
_{B}=\epsilon _{F}\equiv \epsilon $),
\begin{eqnarray}
&&\phi _{B,F}\equiv \sqrt{\frac{2}{\lambda 
}}a_{\mathrm{ho}}^{-1}\tilde{\phi}%
_{B,F},~t\equiv \frac{m}{\hbar }\left( \frac{\lambda }{2\pi }\right) ^{2}%
\tilde{t},~x\equiv \frac{\lambda }{2\pi }\tilde{x},\nonumber \\
&&\epsilon =\tilde{\epsilon}%
\frac{(2\pi \hbar )^{2}}{\lambda ^{2}m}  \label{scaling}
\end{eqnarray}%
Eqs. (\ref{g}) and (\ref{h}) are cast in the following dimensionless form,
\begin{eqnarray}
i\frac{\partial \phi _{B}}{\partial t} &=&-\frac{1}{2}\frac{\partial \phi
_{B}}{\partial x^{2}}+{\ g_{B}}|\phi _{B}|^{2}\phi _{B}  \nonumber \\
&+&{\ g_{BF}}|\phi _{F}|^{2}\phi _{B}-\epsilon \cos \left( 2x\right) \phi
_{B},  \label{q1} \\
i\frac{\partial \phi _{F}}{\partial t} &=&-\frac{1}{2}\frac{\partial \phi
_{F}}{\partial x^{2}}+g_{F}|\phi _{F}|^{4/3}\phi _{F}  \nonumber \\
&+&{g_{BF}}|\phi _{B}|^{2}\phi _{F}-\epsilon \cos \left( 2x\right) \phi _{F},
\label{q2}
\end{eqnarray}%
with $g_{B}\equiv a_{BB}\lambda /(\pi a_{\mathrm{ho}}^{2})$, $%
g_{BF}=a_{BF}\lambda /(\pi a_{\mathrm{ho}}^{2})$ and $g_{F}=(3
/10)(3\lambda ^{2}/4\pi a_{\mathrm{ho}}^{2})^{2/3}$ (tildes are dropped
here), the rescaled 1D wave functions being subject to normalization
conditions
\begin{equation}
\int_{-\infty }^{+\infty }\left\vert \phi _{B,F}(x)\right\vert
^{2}dx=N_{B,F}.  \label{N}
\end{equation}

Stationary solutions to Eqs. (\ref{q1}) and (\ref{q2}) are looked for in the
ordinary form, $\phi _{B,F}=\exp \left( -i\mu _{B,F}t\right) u_{B,F}(x)$,
with real chemical potentials $\mu _{B}$ and $\mu _{F}$, and real functions $%
u_{B,F}(x)$ obeying equations (with $u^{\prime \prime }\equiv d^{2}u/dx^{2}$%
)
\begin{eqnarray}
\mu _{B}u_{B} &=&-\frac{u_{B}^{\prime \prime }}{2}+{\ g_{B}}u_{B}^{3}+{\ 
g_{BF}}%
u_{F}^{2}u_{B}-\epsilon \cos \left( 2x\right) u_{B}, \nonumber \\ 
\label{eq1} \\
\mu _{F}u_{F} &=&-\frac{u_{F}^{\prime \prime }}{2}+g_{F}u_{F}^{7/3}+{\ 
g_{BF}}%
u_{B}^{2}u_{F}-\epsilon \cos \left( 2x\right) u_{F}. \nonumber \\ 
\label{eq2}
\end{eqnarray}%
Our main objective is to construct families of fundamental GS solutions to
Eqs. (\ref{eq1}) and (\ref{eq2}), with $\mu _{B,F}$ belonging to the first
two finite bandgaps in the model's linear spectrum [the bandgaps are induced
by the dimensionless potential, $-\epsilon \cos \left( 2x\right) $, common
to both components].

%%In the above-mentioned case of very different masses of the two Bose and
%%Fermi components, one can use the same rescaling as in Eq. (\ref{scaling}),
%%where $m$ is the smallest atomic mass. Then, in the limit of $%
%%m_{F}/m_{B}\rightarrow 0$ (which may imply a mixture of $^{6}$Li with a
%%heavy bosonic species, and means that $m=m_{F}$ is used in the rescaling
%%relations), the second derivative drops out from Eq. (\ref{eq1}), hence it
%%may be solved to eliminate the boson wave function,%
%%\begin{equation}
%%u_{B}^{2}=\frac{\mu _{B}}{g_{B}}+\frac{\epsilon _{B}}{g_{B}}\cos (2x)-\frac{%
%%g_{BF}}{g_{B}}u_{F}^{2}.  \label{uB1}
%%\end{equation}%
%%The substitution of this expression in Eq. (\ref{eq2}) yields a single
%%equation for the fermionic order parameter:%
%%\begin{equation}
%%\hat{\mu}_{F}u_{F}=-\frac{1}{2}u_{F}^{\prime \prime }+g_{F}u_{F}^{7/3}-\frac{%
%%g_{BF}^{2}}{g_{B}}u_{F}^{3}-\hat{\epsilon}_{F}\cos \left( 2x\right) u_{F},
%%\label{single1}
%%\end{equation}%
%%where $\hat{\mu}_{F}\equiv \mu _{F}-\left( g_{BF}/g_{B}\right) \mu _{B}$,
%%and $\hat{\epsilon}_{F}\equiv \epsilon _{F}-\left( g_{BF}/g_{B}\right)
%%\epsilon _{B}$. Equation (\ref{single1}) features the mixed nonlinearity,
%%with the cubic term being effectively \emph{attractive}.
%%
%%Similarly, in the opposite limit of $m_{B}/m_{F}\rightarrow 0$ (which means $%
%%m=m_{B}$ is used for the rescaling), the second derivative drops out from
%%eq. (\ref{eq2}), making it possible to eliminate the boson order parameter
%%[cf. Eq. (\ref{uB1})].

\section{Variational approximation (VA)}

To derive the VA \cite{VA} for the fundamental GS solutions to stationary
equations (\ref{eq1}) and (\ref{eq2}), we use the obvious fact that these
equations, along with normalization conditions (\ref{N}), can be derived
from Lagrangian
\begin{eqnarray}
L &=&\int_{-\infty }^{+\infty }\left[ \mu _{B}u_{B}^{2}+\mu _{F}u_{F}^{2}-%
\frac{1}{2}\left( u_{B}^{^{\prime }}\right) ^{2}-\frac{1}{2}\left(
u_{F}^{^{\prime }}\right) ^{2}\right.   \nonumber \\
&+&\epsilon \cos (2x)(u_{B}^{2}+u_{F}^{2})-\frac{1}{2}g_{B}u_{B}^{4}-\frac{3%
}{5}g_{F}u_{F}^{10/3}  \nonumber \\
&&\left. -g_{BF}u_{B}^{2}u_{F}^{2}\right] dx-\mu _{B}N_{B}-\mu _{F}N_{F},
\label{L}
\end{eqnarray}%
where $\mu _{B}$ and $\mu _{F}$ play the role of Lagrange multipliers
associated with conditions (\ref{N}). Aiming to find solitons with a compact
profile, we adopt the Gaussian ansatz \cite{VA},
\begin{equation}
u_{B,F}(x)=\pi ^{-1/4}\sqrt{\frac{N_{B,F}\aleph_{B,F}}{w_{B,F}}}\exp 
\left( -%
\frac{x^{2}}{2w_{B,F}^{2}}\right) ,  \label{ansatz}
\end{equation}%
where the variational parameters are soliton's widths $w_{B,F}$ and reduced
norms $\aleph_{B,F}$, as well as chemical potentials $\mu _{B,F}$. The
substitution of the ansatz in Lagrangian (\ref{L}) yields
\begin{eqnarray}
L &=&\mu _{B}N_{B}\aleph_{B}+\mu 
_{F}N_{F}\aleph_{F}-\frac{N_{B}\aleph_{B}}{4w_{B}^{2}}-%
\frac{N_{F}\aleph_{F}}{4w_{F}^{2}}  \nonumber \\
&+&\epsilon N_{B}\aleph_{B}e^{-w_{B}^{2}}+\epsilon 
N_{F}\aleph_{F}e^{-w_{F}^{2}}-\frac{%
g_{B}}{\pi ^{1/2}2^{3/2}}  \nonumber \\
&\times &\frac{N_{B}^{2}\aleph_{B}^{2}}{w_{B}}-\frac{g_{F}}{\pi 
^{1/3}\left(
5/3\right) ^{3/2}}\frac{N_{F}^{5/3}\aleph_{F}^{5/3}}{w_{F}^{2/3}}  
\nonumber \\
&-&\frac{g_{BF}N_{B}N_{F}\aleph_{B}\aleph_{F}}{\sqrt{\pi 
(w_{B}^{2}+w_{F}^{2})}}-\mu
_{B}N_{B}-\mu _{F}N_{F}.  \label{LGauss}
\end{eqnarray}%
The first pair of the variational equations following from Eq. (\ref{LGauss}%
), $\partial L/\partial \mu _{B,N}=0$, yield $\aleph_{B}=\aleph_{F}=1$ 
[which 
implies
that the norms of the two components of ansatz (\ref{ansatz}) are $N_{B}$
and $N_{F}$, in accordance with Eq. (\ref{N})]. Therefore, 
$\aleph_{B}=\aleph_{F}=1$
is substituted in other equations below. The other pair of the variational
equations, $\partial L/\partial w_{B,N}=0$, can be written as
\begin{equation}
1+\frac{g_{B}N_{B}}{(2\pi )^{1/2}}w_{B}+\frac{2g_{BF}N_{F}w_{B}^{4}}{\sqrt{%
\pi }(w_{B}^{2}+w_{F}^{2})^{3/2}}=4\epsilon w_{B}^{4}e^{-w_{B}^{2}},
\label{W1}
\end{equation}%
\begin{equation}
1+\frac{4\sqrt{3}g_{F}N_{F}^{2/3}}{5\sqrt{5}\pi ^{1/3}}w_{F}^{4/3}+\frac{%
2g_{BF}N_{B}w_{F}^{4}}{\sqrt{\pi }(w_{B}^{2}+w_{F}^{2})^{3/2}}=4\epsilon
w_{F}^{4}e^{-w_{F}^{2}}.  \label{W2}
\end{equation}%
The remaining equations, $\partial L/\partial \aleph_{B,F}=0$, yield 
expressions
for chemical potentials $\mu _{B,F}$ of the boson and fermion components of
the fundamental GS as functions of the interaction coefficients, $g_{B}$, $%
g_{F}$,\ and $g_{BF}$:
\begin{equation}
\mu _{B}=\frac{1}{4w_{B}^{2}}+\frac{g_{B}}{\sqrt{2\pi }}\frac{N_{B}}{w_{B}}+%
\frac{g_{BF}N_{F}}{\sqrt{\pi (w_{B}^{2}+w_{F}^{2})}}-\epsilon e^{-w_{B}^{2}},
\label{muG1}
\end{equation}%
\begin{equation}
\mu _{F}=\frac{1}{4w_{F}^{2}}+\frac{\sqrt{3}g_{F}}{\pi ^{1/3}\sqrt{5}}\frac{%
N_{F}^{2/3}}{w_{F}^{2/3}}+\frac{g_{BF}N_{B}}{\sqrt{\pi (w_{B}^{2}+w_{F}^{2})}%
}-\epsilon e^{-w_{F}^{2}}.  \label{muG2}
\end{equation}

In the limit of $g_{B,F}$, $g_{BF}\rightarrow 0$, Eqs. (\ref{W1}) and (\ref%
{W2}) reduce to a single equation,
\begin{equation}
4\epsilon w_{B,F}^{4}\exp \left( -w_{B,F}^{2}\right) =1,  \label{limit}
\end{equation}%
which has real solutions for $\epsilon >e^{2}/8\approx \allowbreak 0.92$.
Formally, solutions of Eq. (\ref{limit}) predict localized solutions to Eqs.
(\ref{eq1}) and (\ref{eq2}) in the linear system, while, as is commonly
known, the linear Schr\"{o}dinger equation with a periodic potential does
not support any localized solution. However, the actual meaning of the limit
case of $g_{B,F}$, $g_{BF}\rightarrow 0$ is that, with solutions to Eq. (\ref%
{limit}) substituted in Eqs. (\ref{muG1}) and (\ref{muG2}), the latter
equations predict a cutoff value, $\left( \mu _{0}\right) _{1}$, for both
chemical potentials, $\mu _{B}$ and $\mu _{F}$, below which fundamental GSs
do not exist, i.e., as a matter of fact, $\left( \mu _{0}\right) _{1}$ is
the \textit{left edge} of the first (lowest) bandgap. A numerically computed
position of the left edge agrees very well with the variational prediction.
For instance, for $\epsilon =5$ (numerical results are displayed below for
this value of $\epsilon $), the VA yields $\left( \mu _{0}\right)
_{1}\approx -2.894$, which almost exactly coincides with its numerically
found counterpart, $\left( \mu _{0}\right) _{1}^{\mathrm{(num)}}\approx
-2.893$.

\section{Numerical results}

\subsection{Families of fundamental gap solitons}

For the numerical solution, Eqs. (\ref{q1}) and (\ref{q2}) were discretized
using the Crank-Nicholson scheme \cite{sk1} and solved (in real time) until
the solution would converge to a stationary wave form that provides a real
solution to Eqs. (\ref{eq1}) and (\ref{eq2}). The numerical solutions was
carried out with time step $0.0005$ and space step $0.025$, in domain $%
-20<x<20$. To find the GSs, the simulations started with an initial
configuration chosen as the ground state, $\phi (x)=(\sqrt{2}c/\pi
)^{1/4}\exp \left( -\sqrt{c/2}x^{2}\right) $, of the quantum-mechanical
linear oscillator with potential $cx^{2}$ (with large $c$, typically $5$ to $%
200$). The OL potential was slowly introduced in the course of numerical
simulation, while the strong harmonic potential kept the
localized state squeezed into a single cell of the OL. After
reaching a stationary bound state in the nonlinear equation
containing the combined OL and harmonic potential, the latter term
was slowly switched off, allowing the establishment of a
well-defined stationary GS (a similar approach to the creation of
GSs in the single-component BEC was proposed in Ref.
\cite{Michal}).

Families of the fundamental GSs predicted by the VA may be
characterized by dependences $g_{B,F}(\mu )$. The same dependences
were found from the numerical solution outlined above. In this
subsection, we display numerical results obtained for compact GSs,
without conspicuous ``tails".

\begin{figure}[tbp]
\begin{center}
{\includegraphics[width=1.\linewidth]{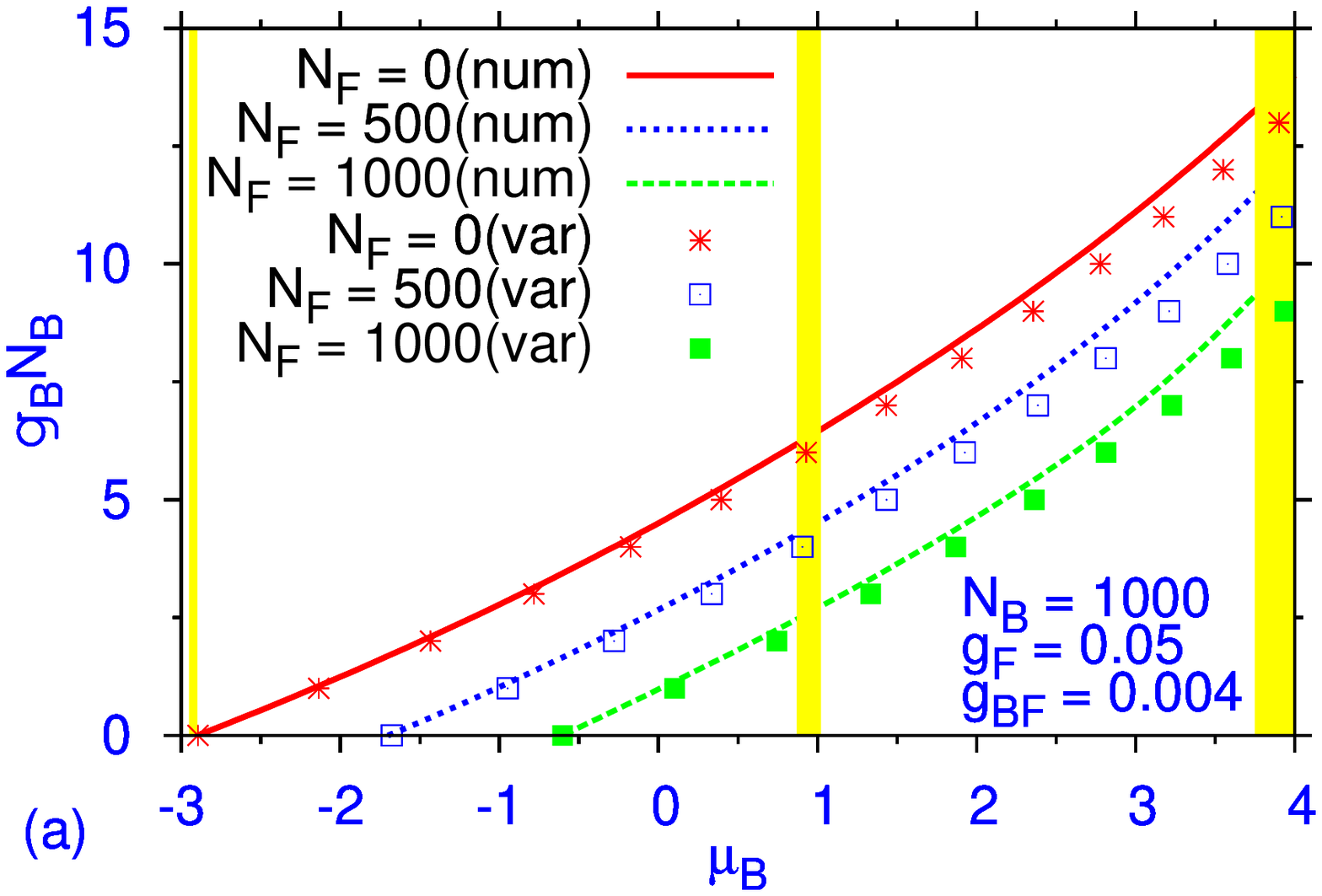}} %\subfigure[] %
{\includegraphics[width=1.\linewidth]{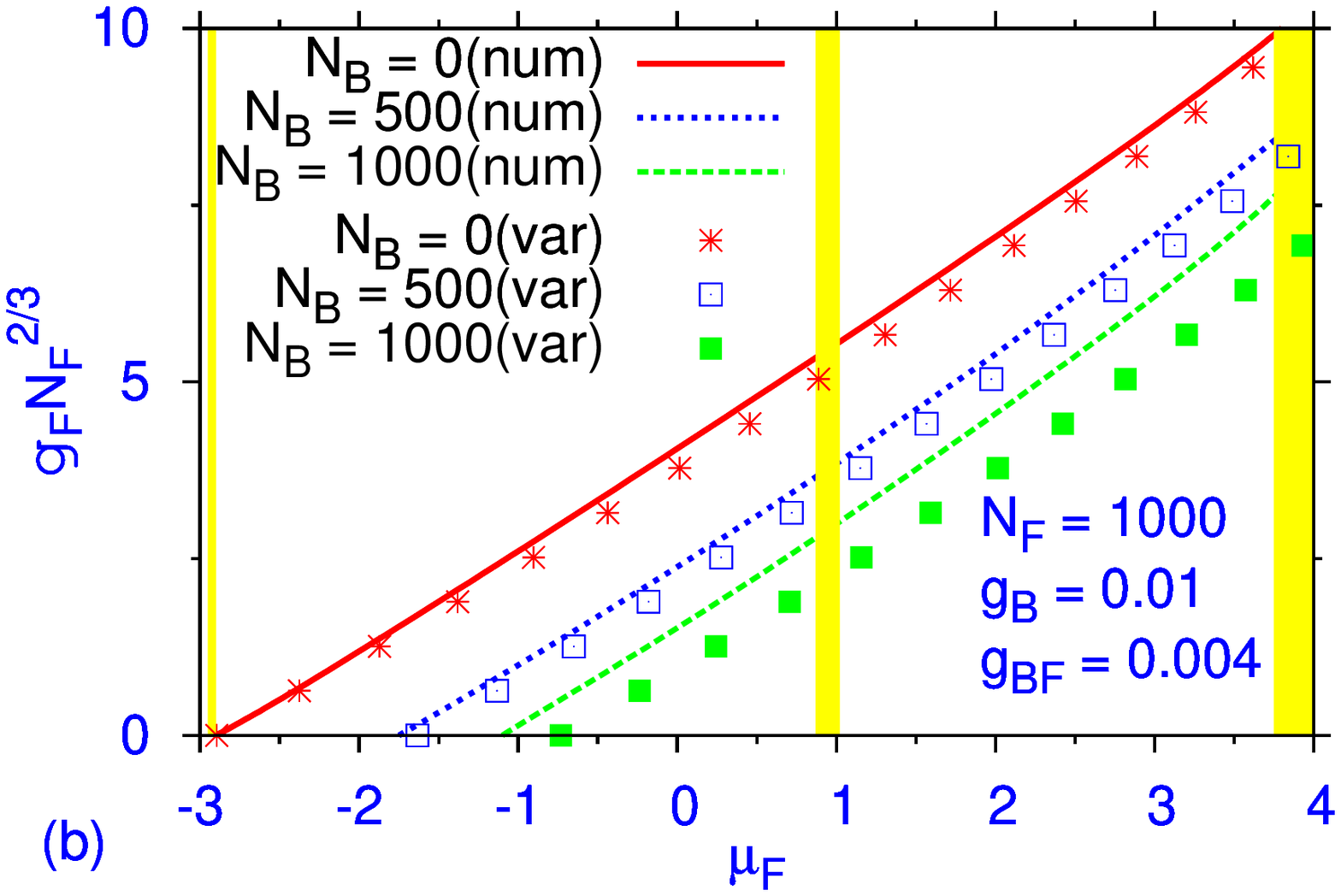}}
\end{center}
\caption{(Color online) The effective nonlinearity coefficients, $g_{B}N_{B}$
for bosons (a) and $g_{F}N_{F}^{2/3}$ for fermions (b), versus the chemical
potentials of the boson and fermion components, $\protect\mu _{B}$ and $%
\protect\mu _{F}$, respectively, for families of the fundamental compact gap
solitons in the first and second bandgaps of periodic potential $\protect%
\epsilon \cos (2x)$ (numerical results are presented in this and
other figures for $\protect\epsilon =5$). As indicated in the
figure, continuous curves and chains of symbols display,
respectively, results obtained from the numerical solution
(``num") and variational approximation (``var"). Shaded vertical
areas represent Bloch bands which separate the gaps.} \label{Fig1}
\end{figure}

The effective nonlinearities, $g_{B}N_{B}$ and $g_{F}N_{F}^{2/3}$, for the
boson and fermion components of the GS families are shown, against the
corresponding chemical potentials, in Figs. \ref{Fig1}(a) and (b), fixing $%
\epsilon =5$ and $g_{BF}=0.004$. In panel (a), we also fix $g_{F}=0.05$ and $%
N_{B}=1000$, while the bosonic-nonlinearity coefficient, $g_{B}$, varies
from $0$ to $0.015$, and the results are displayed for different fixed
values of $N_{F}$ (in particular, $N_{F}=0$ corresponds to the ordinary GSs
in the self-repulsive BEC). The families of variational solutions displayed
in Fig. \ref{Fig1}(a) commence at $\mu \approx -2.894$ (as said above, it
almost exactly coincides with the actual left edge of the first bandgap). In
panel (b) of Fig. \ref{Fig1}, we fix $g_{B}=0.01$ and $N_{F}=0.05$, while
the fermion-nonlinearity coefficient, $g_{F}$, varies from $0$ to $0.1$, the
results being shown for several different fixed values of $N_{B}$ (the case
of $N_{B}=0$ corresponds to pure fermionic GSs in the BCS superfluid). The
agreement between the VA and numerical results is very good, with a caveat
that the VA ignores the presence of the Bloch band separating the two gaps
(i.e., the VA-generated solution branches extend across the band).

\begin{figure}[tbp]
\begin{center}
{\includegraphics[width=.7\linewidth]{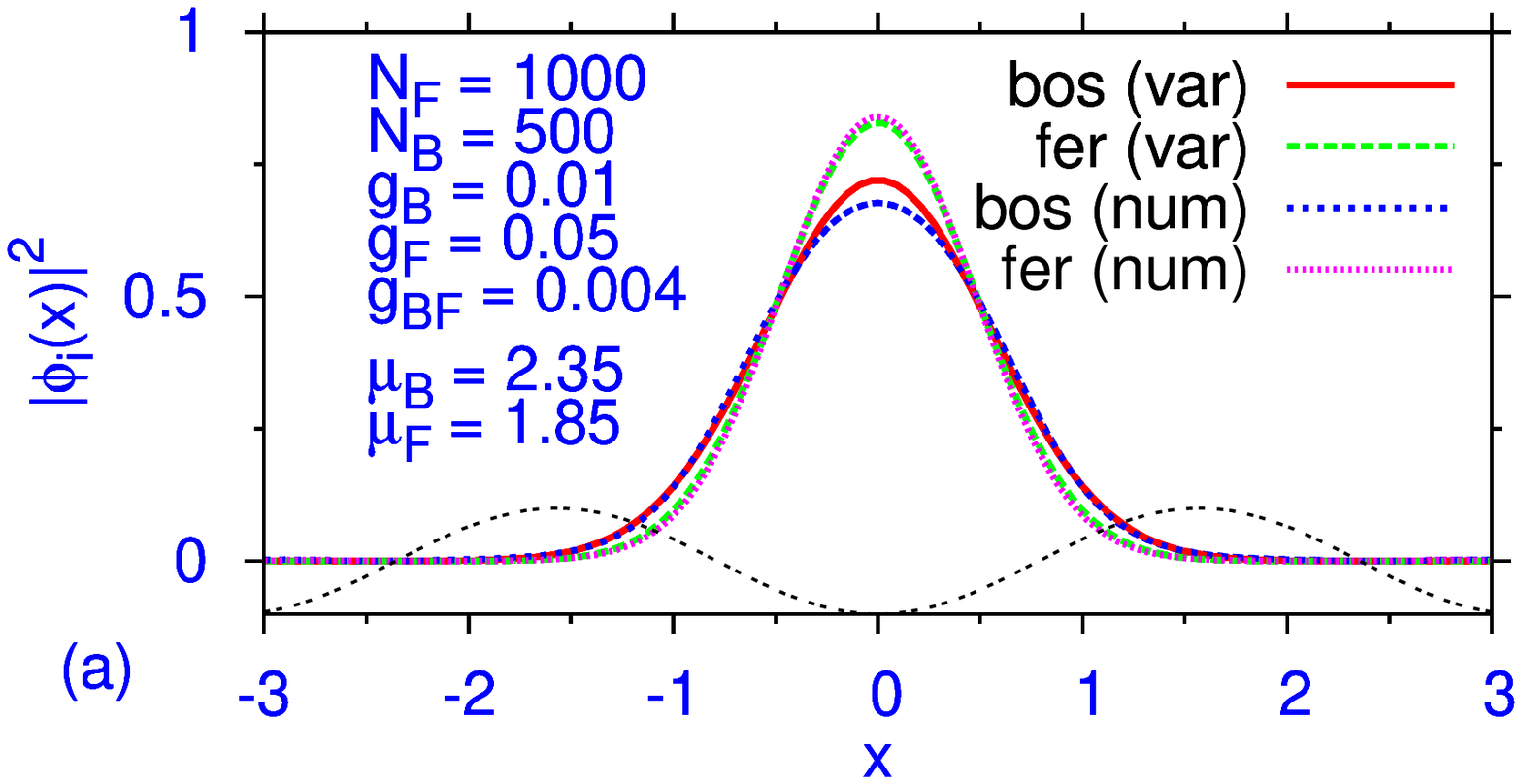}} %\subfigure[] %
{\includegraphics[width=.7\linewidth]{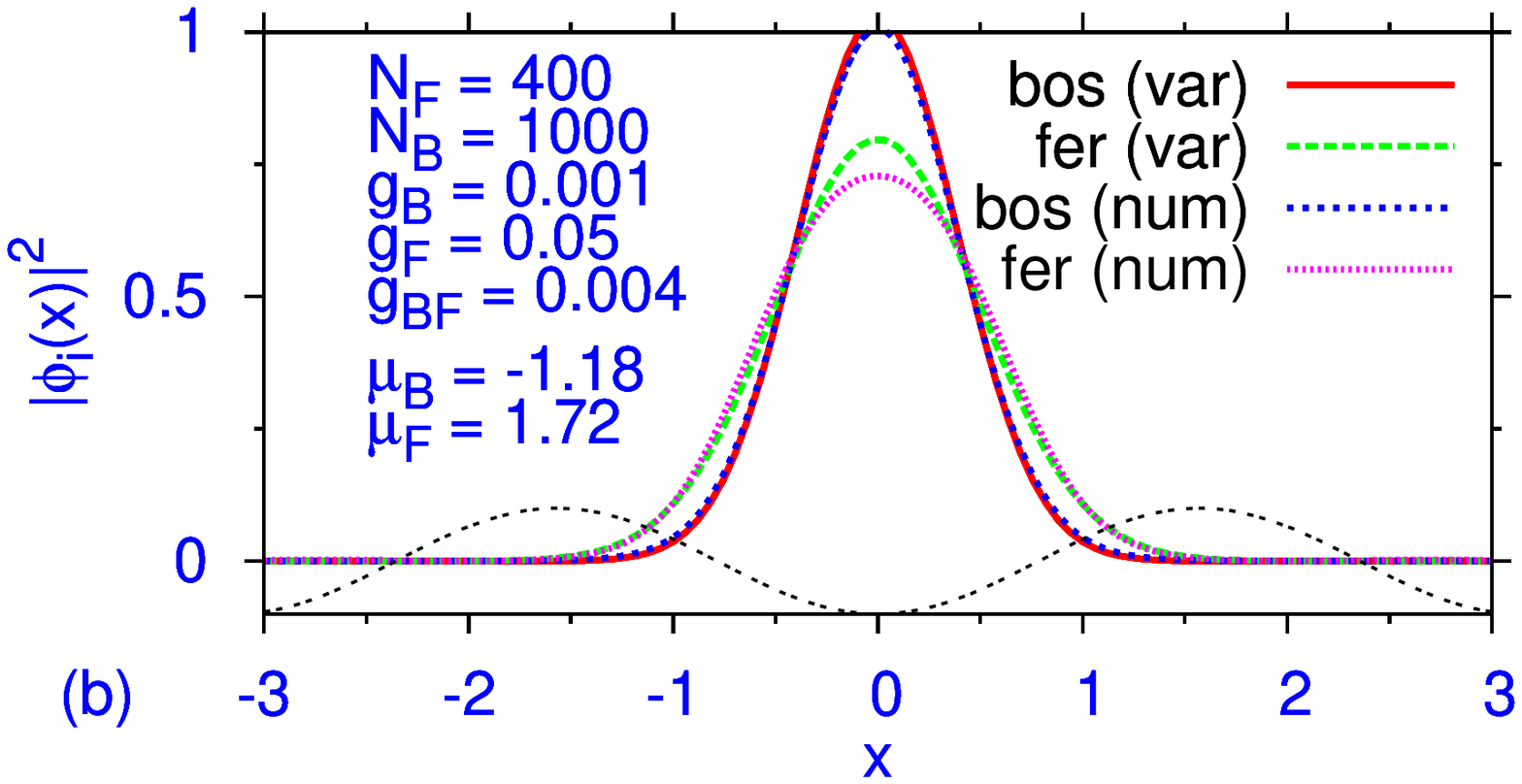}} {\includegraphics[width=.7%
\linewidth]{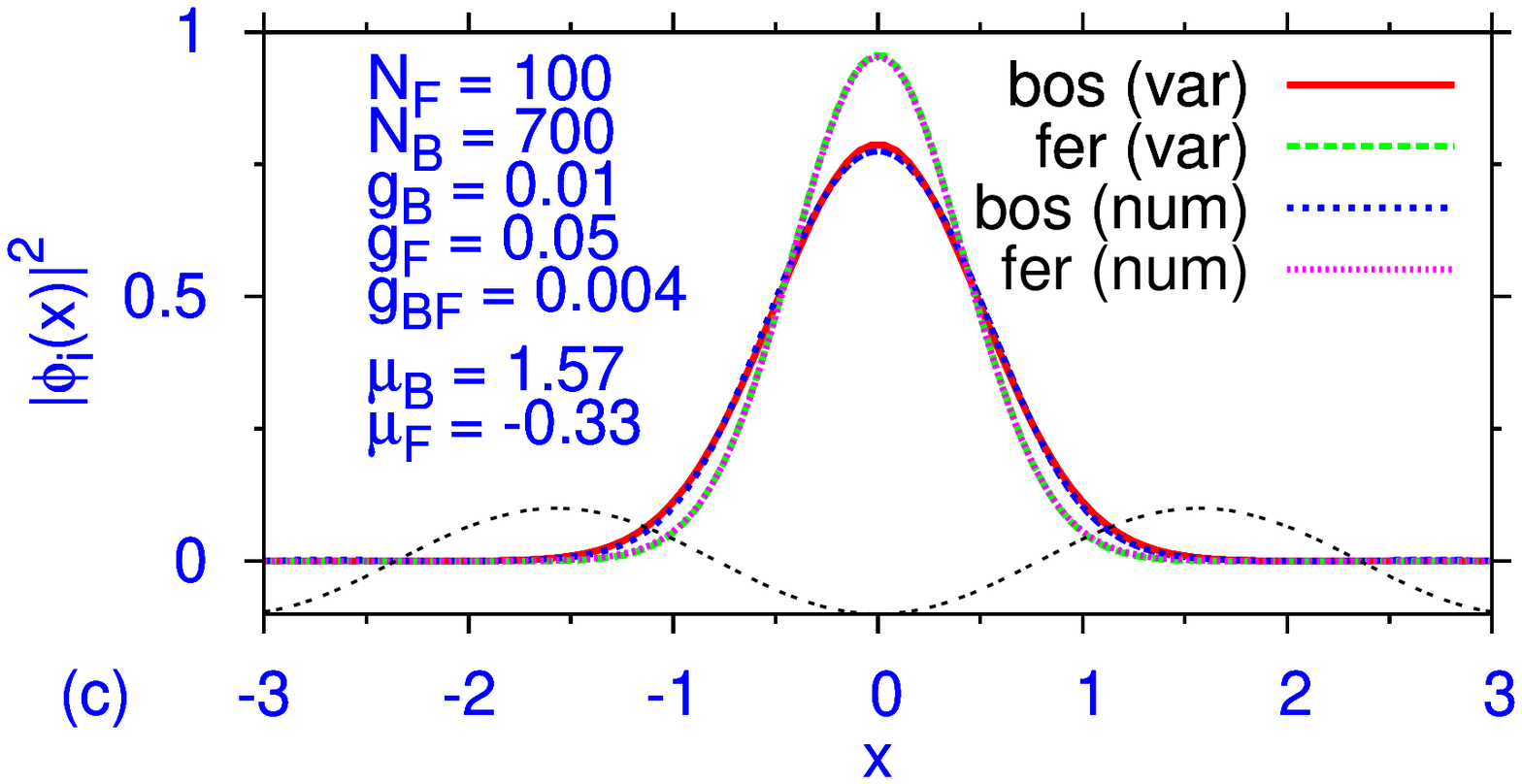}} {\includegraphics[width=.7\linewidth]{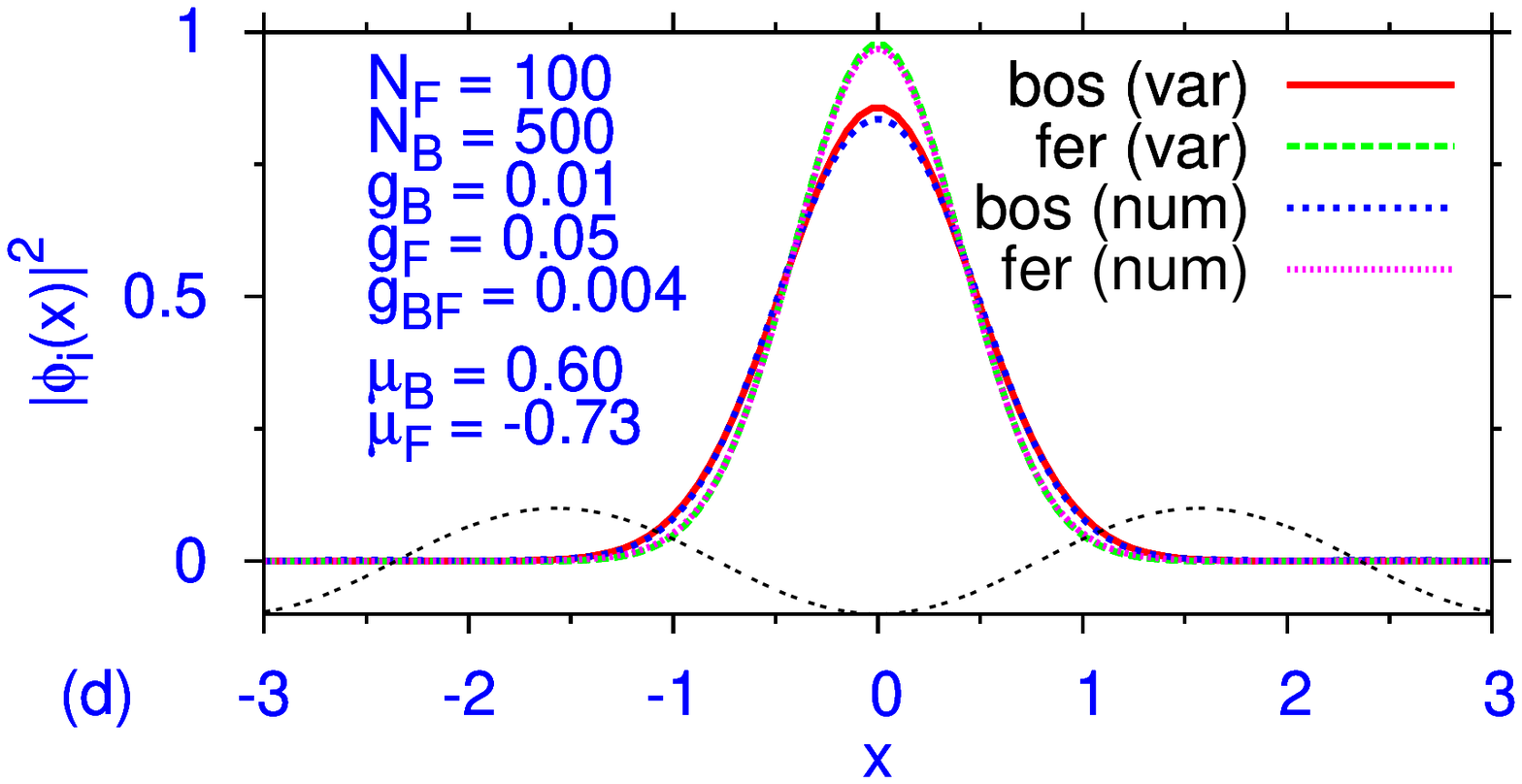}}
\end{center}
\caption{(Color online) Profiles of the probability density, $|\protect\phi %
_{B,F}(x)|^{2}$ [subject to normalization $\protect\int_{\infty }^{\infty }|%
\protect\phi _{B,F}(x)|^{2}=1$] in the boson (``bos") and fermion
(``fer") components of the fundamental gap solitons, in four
different cases with (a) both components of the soliton having
their chemical potentials, $\protect\mu _{B}$ and $\protect\mu
_{F}$, in the second bandgap, (b) $\protect\mu _{B}$ and
$\protect\mu _{F}$ in the first and second gaps, respectively; (c)
vice versa to (b); (d) both components the first bandgap. In terms
of Ref. \protect\cite{Arik}, the two-component gap solitons in
panels (a) and (d) are of the\textit{\ intra-gap} type, while the
ones in (b) and (c) are\textit{\ inter-gap} solitons. The thin
dotted lines, here and in other figures, represent the OL
potential, $\propto -\cos (2x)$.} \label{Fig2}
\end{figure}

In Fig. \ref{Fig2}, typical shapes of the numerically found GSs are compared
with their counterparts predicted by the VA at the following sets of
parameter values: (a) $N_{F}=1000,N_{B}=500,g_{B}=0.01$; (b) $%
N_{F}=200,N_{B}=1000,g_{B}=0.001$; (c) $N_{F}=100,N_{B}=700,g_{B}=0.01$; (d)
$N_{F}=100,N_{B}=500,g_{B}=0.01$. In all the four cases, $g_{F}=0.05$ and $%
g_{BF}=0.004$ are fixed. As written in the caption to Fig. \ref{Fig2}, these
parameters are chosen to represent four different cases, as concerns the
location of the two chemical potentials, $\mu _{B}$ and $\mu _{F}$, in the
two lowest bandgaps of the OL-induced linear spectrum: panels (a) and (d)
represent the two-component solitons of the intra-gap type, while (b) and
(c) represent inter-gap solitons, in terms of Ref. \cite{Arik}. The
agreement between the numerical and variational results is, again, quite
noteworthy.

\begin{figure}[tbp]
\begin{center}
{\includegraphics[width=.7\linewidth]{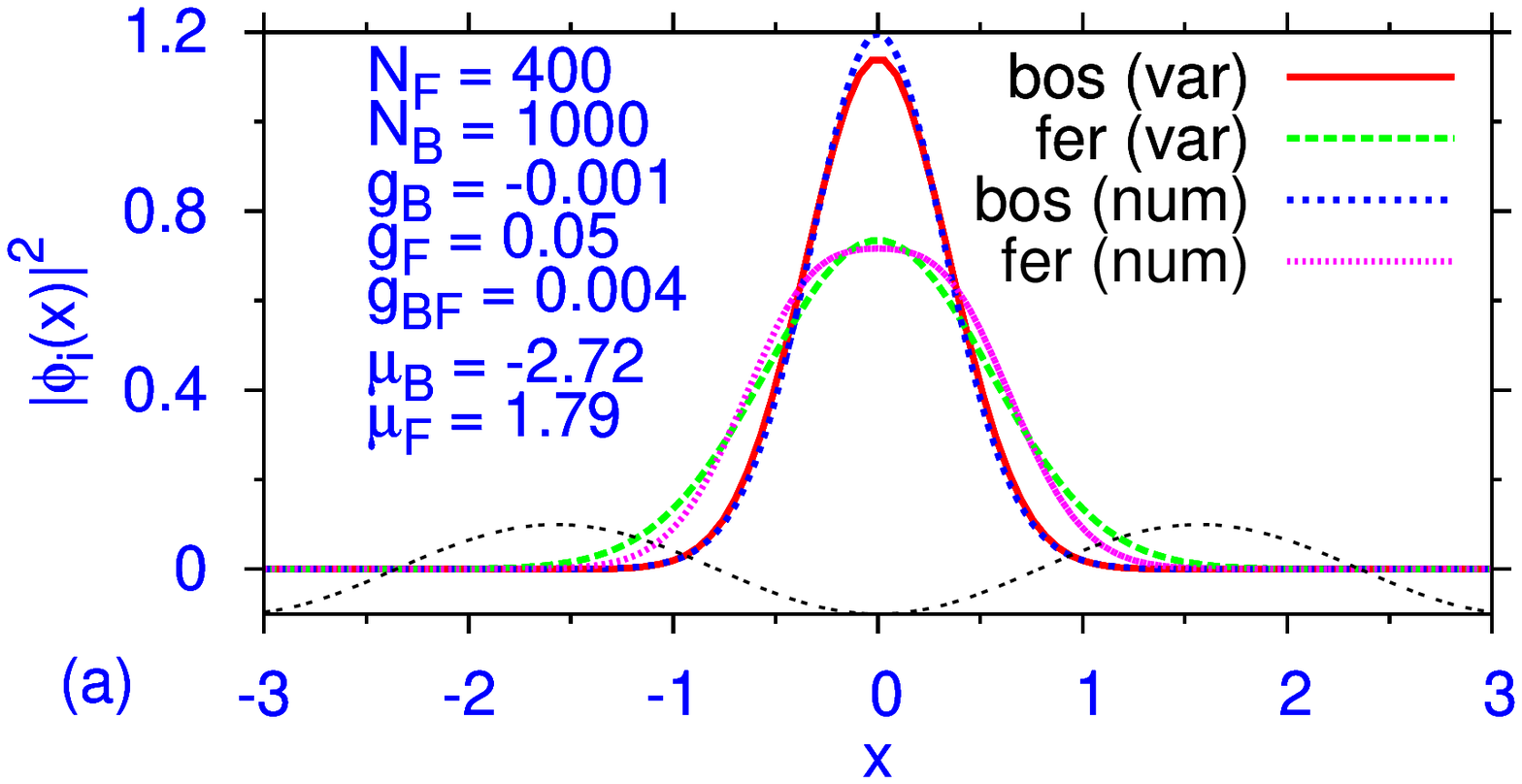}} %\subfigure[] %
{\includegraphics[width=.7\linewidth]{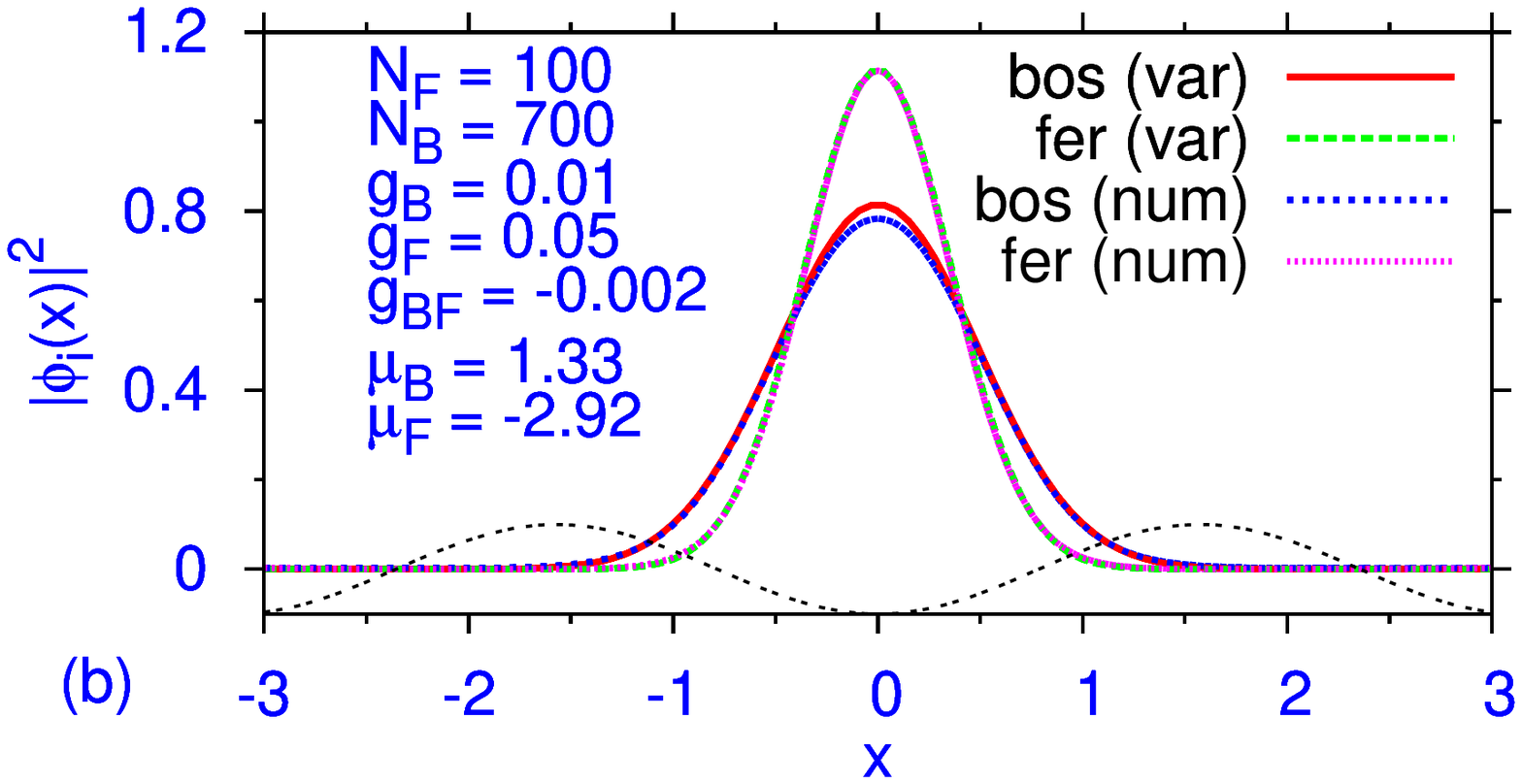}} 
\end{center}
\caption{(Color online) Same as in Fig.  \ref{Fig2} with (a) 
intraspecies 
attraction 
among bosonic atoms corresponding to a negative $g_B$
and (b) interspecies attraction between bosons 
and 
fermions corresponding to a negative $g_{BF}$. 
In (a) the bosonic GS is in the first bandgap and fermionic 
GS is in the second bandgap, in (b) vice versa to (a). }
\label{FigX}
\end{figure}

In Fig. \ref{Fig2} all the nonlinearities are taken as positive 
(repulsive). 
Another problem of interest is to investigate if GSs in the first two 
bandgaps can be  found in 
both components  
when one of the nonlinearities ($g_B$ or 
$g_{BF}$)
becomes 
attractive. (We recall that when $g_{BF}$ is sufficiently attractive, 
bright Bose and Fermi solitons can be generated without the OL 
potential \cite{Sadhan-BFsoliton}. Such bright solitons will not have 
the chemical potential in 
the  first two bandgaps and hence are not GSs.)  We find that GSs 
in the first two bandgaps can indeed be found when either $g_B$ or
$g_{BF}$ turns negative. We show typical results for this case in Fig. 
\ref{FigX} (a) and (b), where we illustrate that by adjusting the 
parameters such GSs can be placed in the first or the second bandgap.
These are {\it inter-gap} solitons. In addition {\it intra-gap} solitons 
can also be created in this case (not illustrated in Fig. \ref{FigX}.)   
The 
solitons shown in Fig. \ref{FigX} are also stable, despite the fact that one 
nonlinearity  coefficient changed its sign to an attractive value. 

\subsection{Nonfundamental solitons}

\begin{figure}[tbp]
\begin{center}
%\subfigure[]
{\includegraphics[width=.8\linewidth]{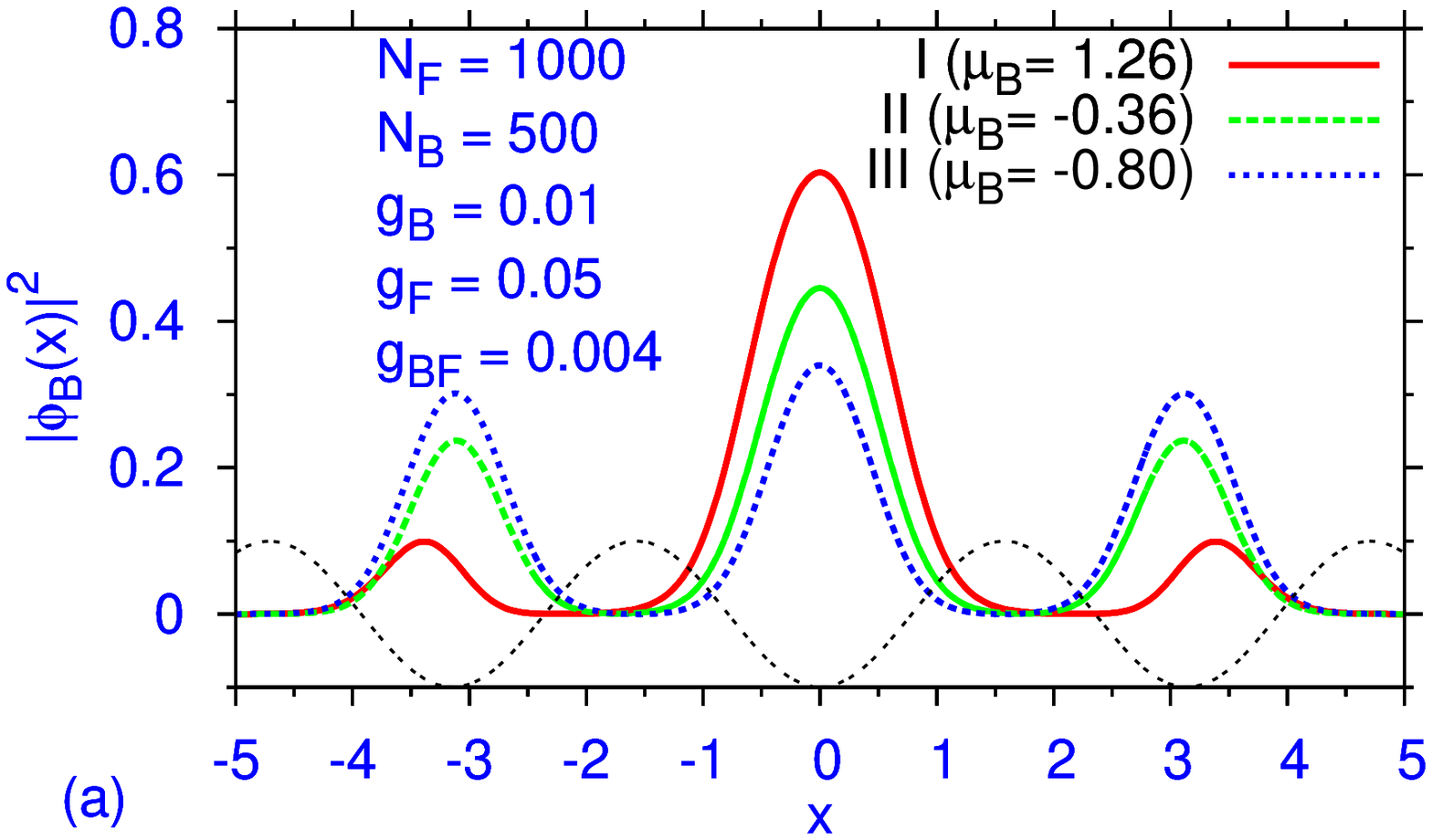}} %\subfigure[] %
{\includegraphics[width=.8\linewidth]{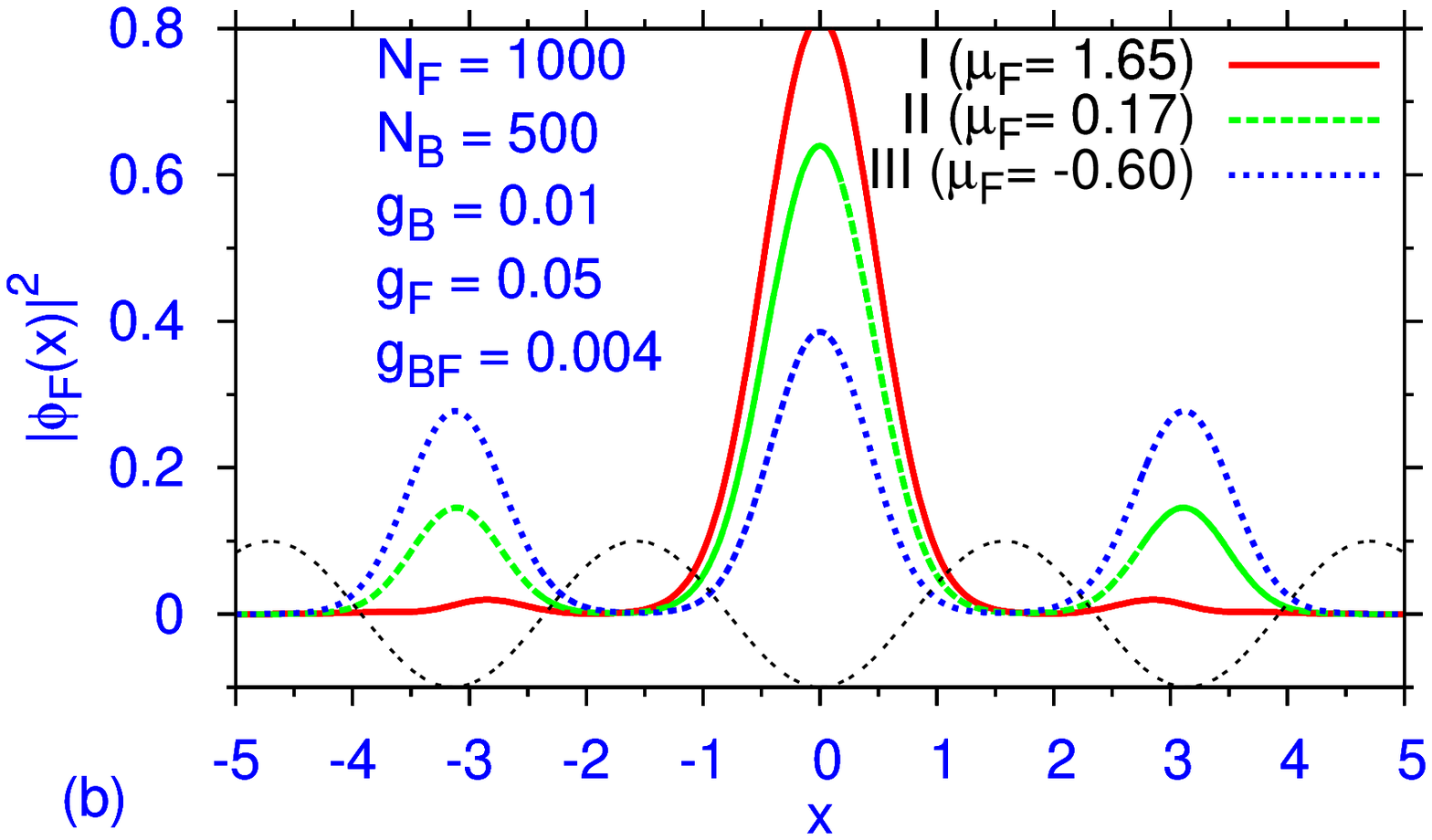}}
\end{center}
\caption{(Color online) Profiles labeled I, II, III [normalized according to
$\protect\int_{\infty }^{\infty }|\protect\phi _{B,F}(x)|^{2}=1$] represent
the bosonic (a) and fermionic (b) components of three distinct stable
\textit{nonfundamental} gap solitons, extended over several lattice cells,
for the same parameters (except for $\protect\mu _{B,F}$) as in Fig. \protect
\ref{Fig2}(a). The corresponding values of the chemical potentials are
indicated in the panels.}
\label{Fig3}
\end{figure}

The fundamental GSs presented by Figs. \ref{Fig1} and \ref{Fig2} are compact
objects with a nearly-Gaussian shape, trapped in a single cell of the OL. On
the other hand, it is well known \cite{GSprediction} that the GSs in
self-repulsive repulsive BEC frequently demonstrate structures extended over
many cells. Stable \textit{nonfundamental} solitons featuring similar
profiles can be found in the present model too, see examples in Fig. \ref%
{Fig3}. They may be considered as bound complexes built of a central GS and
additional pulses located in adjacent cells of the OL. Similar to the
patterns shown Fig. \ref{Fig3}, which extend over three lattice cells, at
smaller values of the chemical potentials it is possible to find dynamically
stable localized structures which are stretched still wider.

\subsection{Stability}

\begin{figure}[tbp]
\begin{center}
%\subfigure[]
{\includegraphics[width=.48\linewidth]{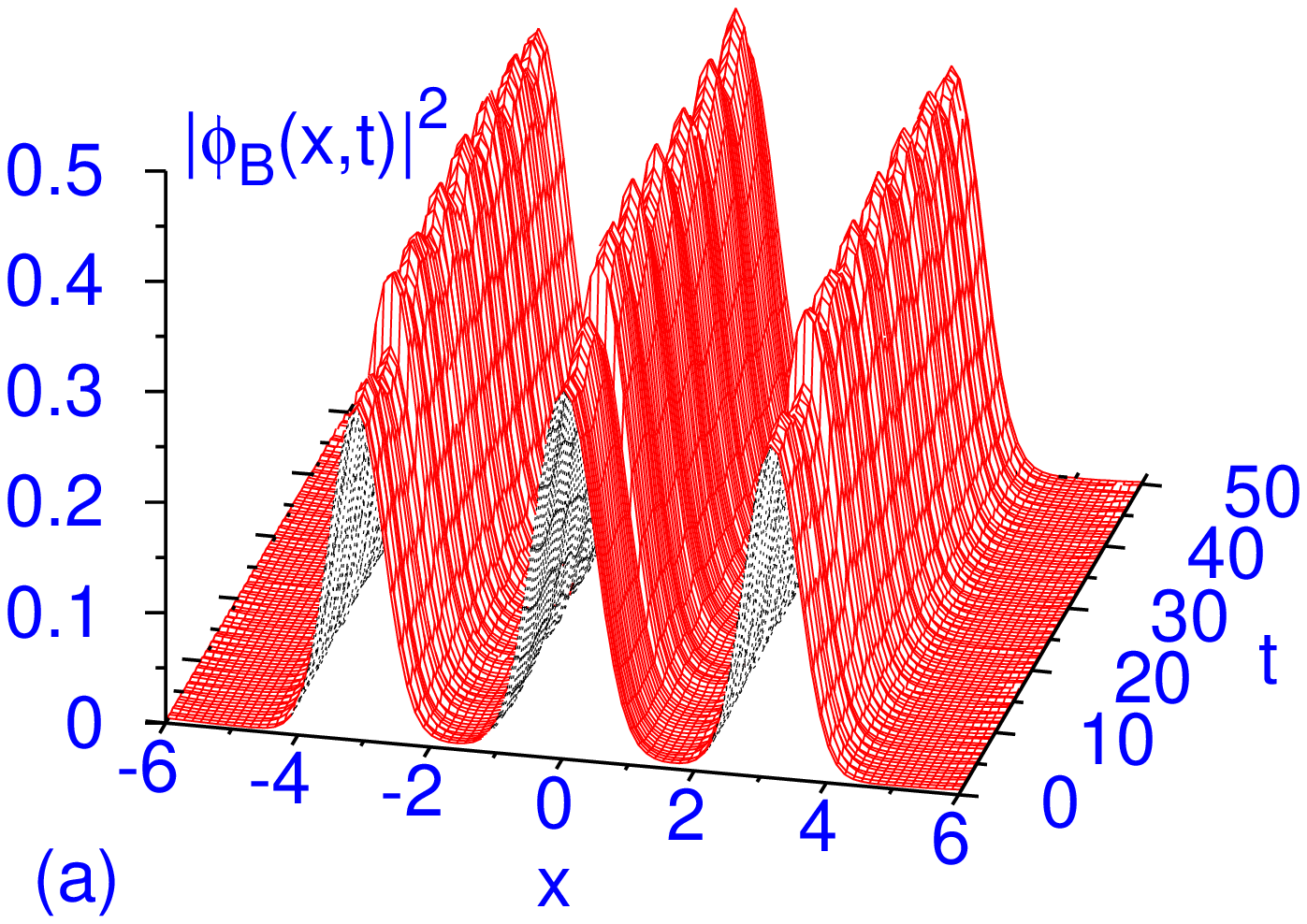}} %\subfigure[] %
{\includegraphics[width=.48\linewidth]{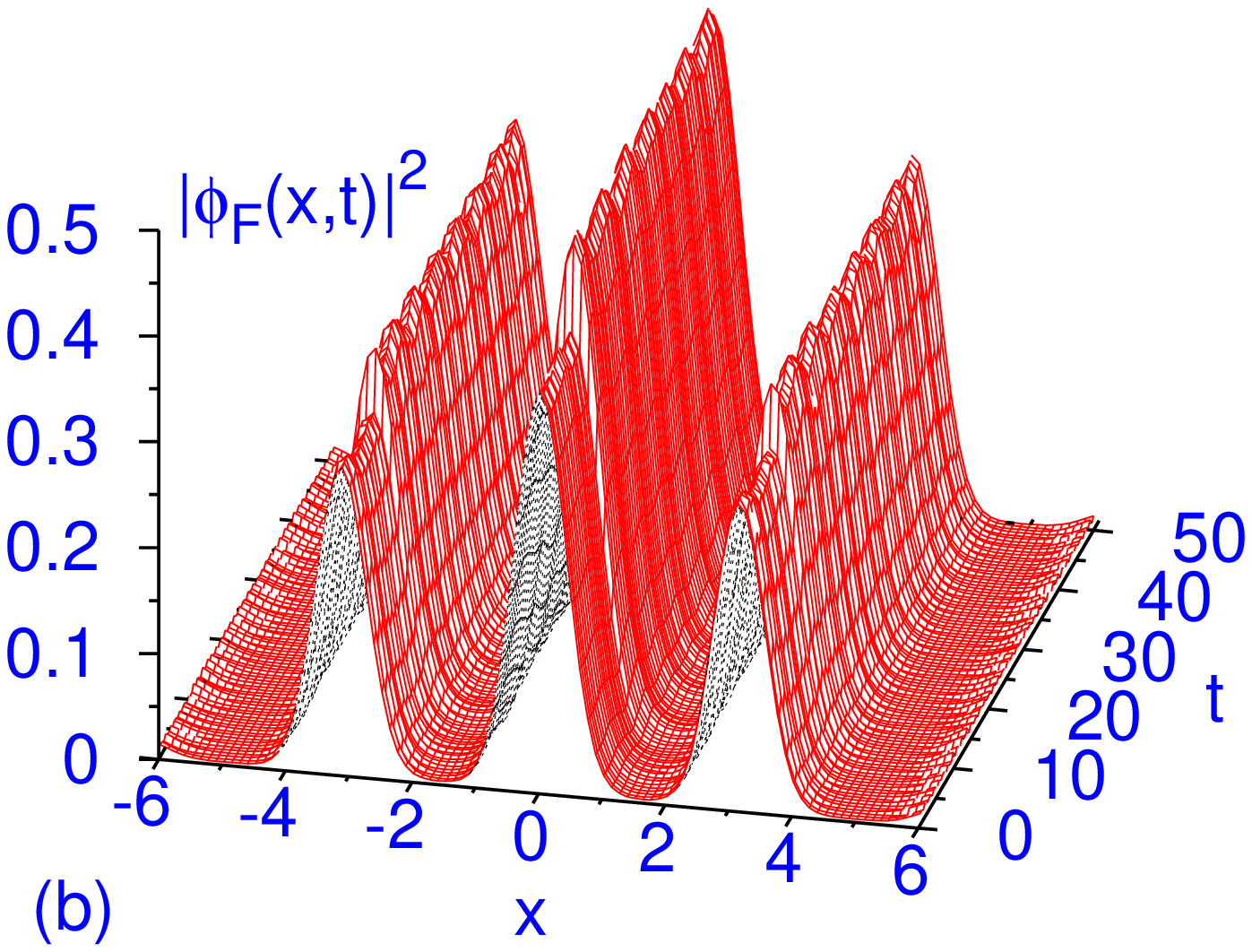}}
\end{center}
\caption{(Color online) The evolution of the boson (a) and fermion (b)
components of the nonfundamental gap soliton, labeled III in Figs. \protect
\ref{Fig3}, with $\protect\mu _{B}=-0.80$ and $\protect\mu _{F}=-0.60$. A
rather strong perturbation was applied to the soliton at $t=10$, in the form
of $\protect\phi _{B,F}(x)\rightarrow 1.1\protect\phi _{B,F}(x)$. The
perturbed soliton remains stable as long as the simulation was running.}
\label{Fig4}
\end{figure}

The way the fundamental and nonfundamental GSs have been found, as a limit
form to which solutions of underlying dynamical equations (\ref{q1}) and (%
\ref{q2}) relax at large values of $t$, strongly suggests that all these
solutions are stable. Their stability was additionally tested by adding
strong perturbations to them. It was thus verified that the entire families
of the GSs presented above are stable, as well as the nonfundamental
solitons displayed in Fig. \ref{Fig3}. For example, the nonfundamental
soliton labeled III in Fig. \ref{Fig3}, after it was obtained as a steady
state to which the evolution of a certain initial configuration converged,
was suddenly multiplied by a perturbing factor, $\phi _{B,F}(x)\rightarrow
1.1\phi _{B,F}(x)$. The perturbed solution demonstrates oscillations without
any onset of instability, as shown in Fig. \ref{Fig4}.

\begin{figure}[tbp]
\begin{center}
%\subfigure[]
{\includegraphics[width=.48\linewidth]{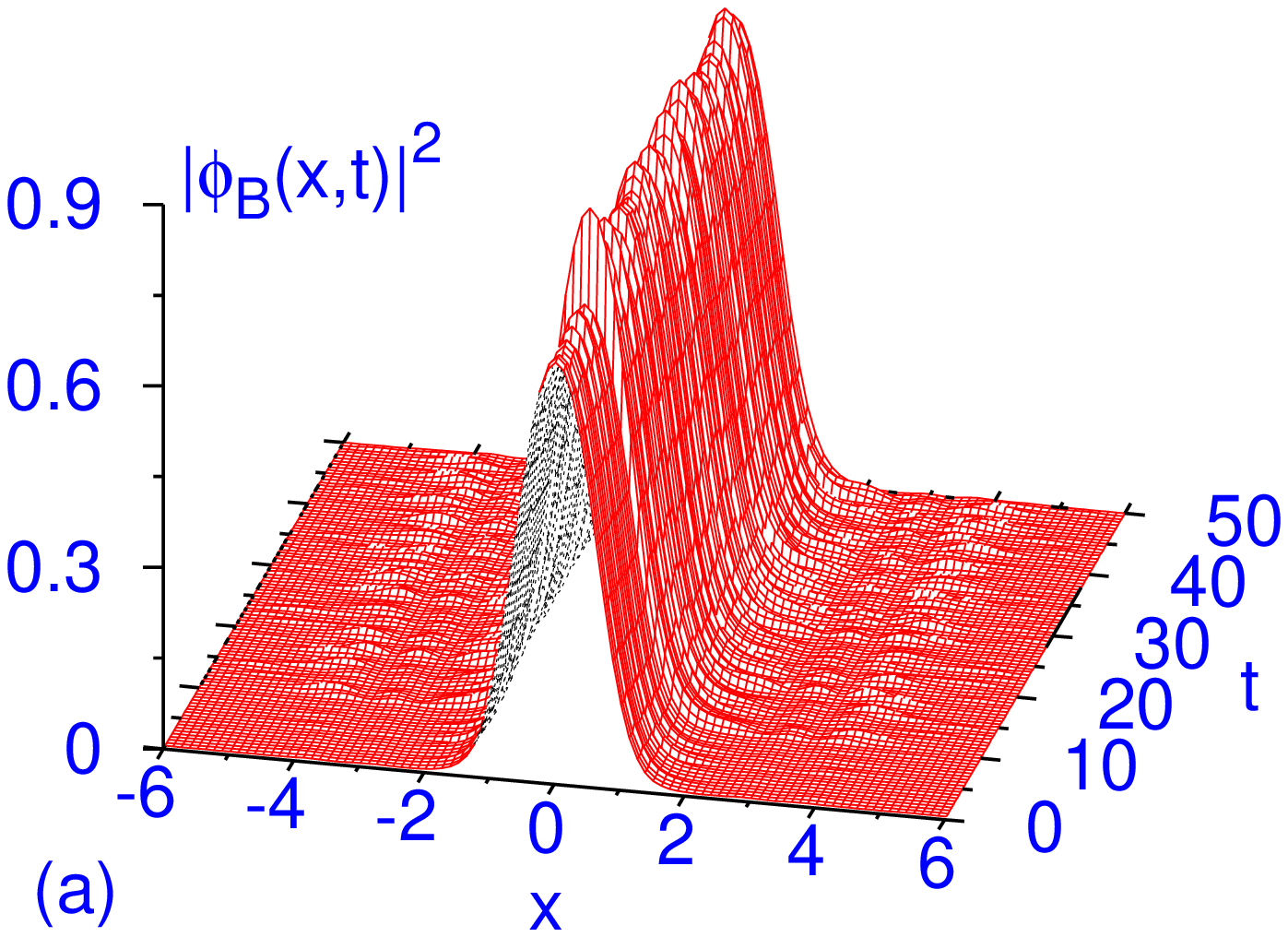}} %\subfigure[] %
{\includegraphics[width=.48\linewidth]{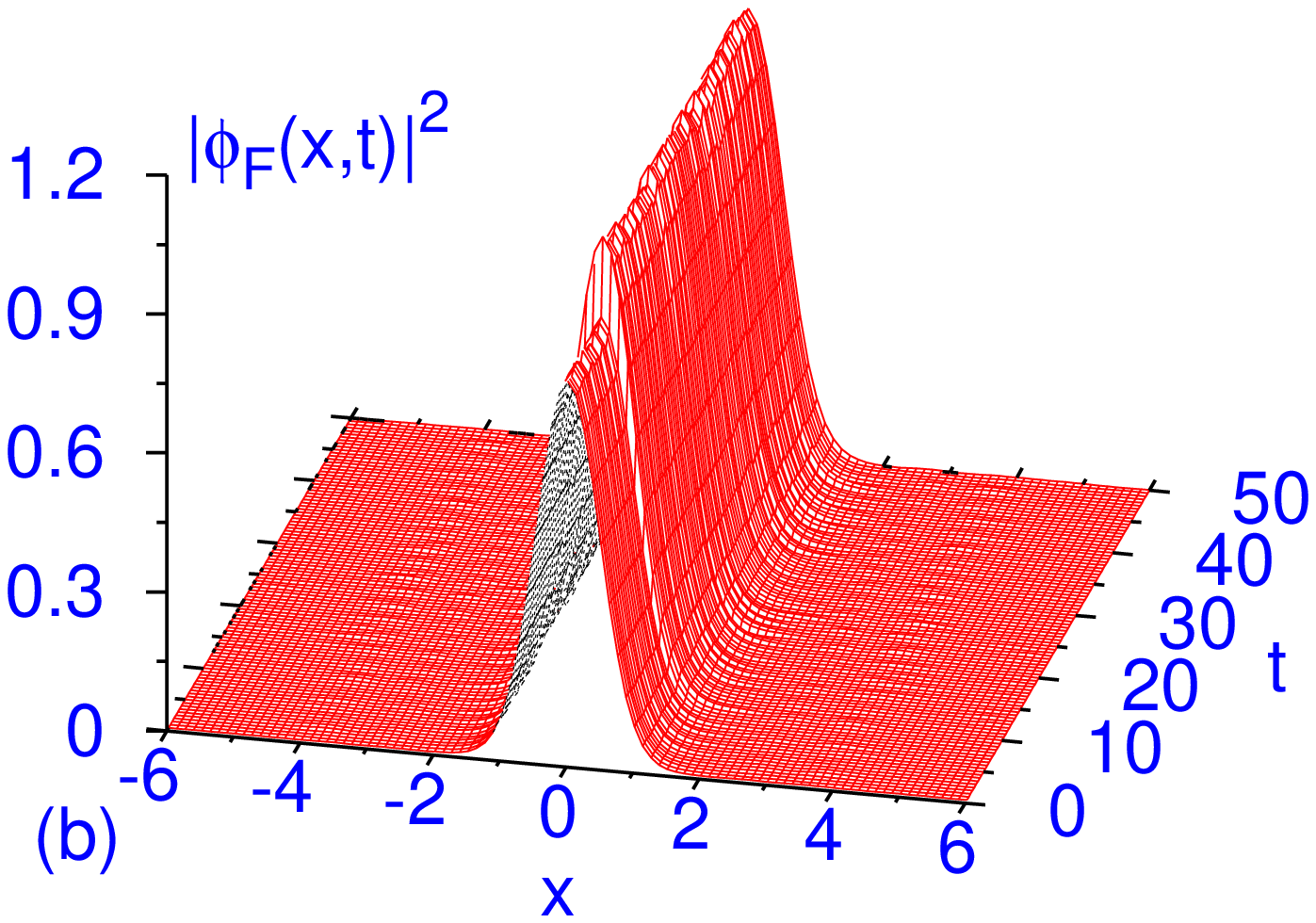}}
\end{center}
\par
\label{Fig5}
\caption{(Color online) The same stability test as shown in Fig. \protect\ref%
{Fig4} for the fundamental soliton from Fig. \protect\ref{Fig2}.}
\end{figure}

A typical example of the stability test for fundamental GSs is
presented in Fig. \ref{Fig5}, for the soliton shown in Fig.
\ref{Fig2}(a). At $t=10$, it was also ``shaken", multiplying it by
factor $1.1$. The subsequent evolution makes the absence of any
instability evident.

\section{Conclusion}

We have studied the possibility of generation of one-dimensional GSs (gap
solitons) in a Bose-Fermi mixture trapped in a periodic OL potential,
assuming that the bosons are in the BEC phase, and fermions are condensed
into the BCS superfluid. A dynamical model of the mixture was derived from
the system's Lagrangian. The resultant coupled equations feature the cubic
nonlinearity in the boson component, and the nonlinear term of power 7/3 in
the equation for the fermion order parameter, with a cubic coupling between
the equations, all the nonlinearities being repulsive.

Families of solutions for compact GSs, trapped, essentially, in the single
lattice cell, were obtained by means of the VA (variational approximation)
based on the Gaussian ansatz, and in the numerical form, in the two lowest
bandgaps of the OL-induced spectrum. For a wide range of values of the
interaction strengths and numbers of boson and fermions in the mixture, the
VA predictions for the GS profiles and their chemical potentials agree very
well with the respective numerical results. The GS families include both
intra- and inter-gap solitons, with the chemical potentials of the boson and
fermion components falling, respectively, in the same or different bandgaps.
In addition to the fundamental GSs, we have also found (in the numerical
form) nonfundamental localized structures, extended over several lattice
sites. The entire families of the fundamental GSs, as well as the
nonfundamental extended states, were checked to be robust against strong
perturbations.

This work can be naturally extended in various directions. One
possibility is to look for GSs in 2D and 3D models of the same type,
study the possibility of the existence and stability of the
two-component boson-fermion gap solitons and investigate their
properties. Another perspective direction would be to study gap solitons
in a boson-fermion mixture with mixed repulsive-attractive interactions
in a systematic form, while in this paper we gave only a few examples.
One can also seek for mixed bound states, in which the boson component
belongs to the semi-infinite gap, while the fermionic chemical potential
still belongs to a finite bandgap and vice versa. Lastly, it may happen
that the coefficients which couple the bosons and fermions to the OL
have opposite signs (one species is red-detuned, while the other one
feels blue detuning), that may give rise to nontrivial stability limits
for two-components GSs.

Here we used a set of mean-field equations for the study of GSs in
Bose-Fermi mixtures.  A rigorous treatment of this problem might be
performed using a many-body Slater-determinant wave function
\cite{Mario}, as, for example, in the case of many-electron scattering
\cite{ps}.

\acknowledgments The work of S.K.A. was supported in part by the FAPESP and
CNPq of Brazil. The work of B.A.M. was supported, in a part, by the Israel
Science Foundation through the Center-of-Excellence grant No. 8006/03, and
the German-Israel Foundation (GIF) through grant No. 149/2006. This author
appreciates the hospitality of Instituto de F\'{\i}sica Te\'{o}rica at the S%
\~{a}o Paulo State University (S\~{a}o Paulo, Brazil).

\end{document}